\documentclass[sigconf]{acmart}
\setcopyright{acmcopyright}
\copyrightyear{2023}
\acmYear{2023}
\acmDOI{}

\acmConference[TheWebConf '23]{The Web Conference}{April 30 - May 04,
  2023}{Austin, TX}
\acmPrice{}
\acmISBN{}

\def\code#1{\texttt{#1}}

\begin{document}

\title{Social network modeling and applications, a tutorial}

\author{Lisette Espin-Noboa}
\email{espin@csh.ac.at}
\affiliation{%
  \institution{CSH \& CEU}
  \country{}
}

\author{Tiago Peixoto}
\email{peixotot@ceu.edu}
\affiliation{%
  \institution{CEU}
  \country{}}

\author{Fariba Karimi}
\email{karimi@csh.ac.at}
\affiliation{%
  \institution{CSH}
  \country{}}
  
\renewcommand{\shortauthors}{Espin-Noboa, Peixoto, and Karimi, 2023}

\begin{abstract}
Social networks have been widely studied over the last century from multiple disciplines to understand societal issues such as inequality in employment rates, managerial performance, and epidemic spread.
Today, these and many more issues can be studied at global scale thanks to the digital footprints that we generate when browsing the Web or using social media platforms.
Unfortunately, scientists often struggle to access to such data primarily because it is proprietary, and even when it is shared with privacy guarantees, such data is either no representative or too big.
In this tutorial, we will discuss recent advances and future directions in \textit{network modeling}.
In particular, we focus on how to exploit synthetic networks to study real-world problems such as data privacy, spreading dynamics, algorithmic bias, and ranking inequalities. 
We start by reviewing different types of generative models for social networks including node-attributed and scale-free networks. Then, we showcase how to perform a network selection analysis to characterize the mechanisms of edge formation of any given real-world network.
\end{abstract}



\maketitle

\section{Target Audience}
This tutorial is targeted to researchers who want to learn more about (i) random network generator models, (ii) deployment of synthetic networks with and without attributes and specific edge formation mechanisms, (iii) model selection given an empirical network, and (iv) how to exploit synthetic networks for sharing data with privacy guarantees, and to understand real-world issues including spreading dynamics, algorithmic bias, and ranking inequalities.
Participants must have a basic knowledge of coding, preferable in Python.\section{Tutors}

\textit{Lisette Espin-Noboa}\footnote{\url{https://www.lisetteespin.info}} is a PostDoc at Central European University (CEU) and at the Complexity Science Hub Vienna (CSH). Her research interests lie at the intersection between computational social science, network science, and AI for social good. She is particularly focused on understanding how edges form in social networks~\cite{espin2017janus}, and how these mechanisms of edge formation may affect machine learning algorithms~\cite{espin2018towards, espin2021explaining, espin2022inequality}, and human behavior\cite{espin2016discovering, espin2019hoprank, walk2017users}.

\textit{Tiago Peixoto}\footnote{\url{https://skewed.de}} is an Associate Professor in the Department of Network and Data Science at the Central European University (CEU). %
His research focuses on the development of methods to extract scientific understanding from network data, as well as the mathematical modeling of network behavior and evolution. He is particularly interested in problems of network inference, where meaningful structural and functional patterns are missing or cannot be obtained by direct inspection or low-order statistics, and require instead more sophisticated approaches based on large-scale generative models and efficient algorithms derived from them~\cite{peixoto2022implicit, peixoto2022disentangling, young2021hypergraph, zhang2020statistical}. 
Many of the methods developed in his work are made available as part of the \code{graph-tool} library~\cite{peixoto_graph-tool_2014}, which is extensively documented.

\textit{Fariba Karimi}\footnote{\url{https://networkinequality.com/people/fariba-karimi}} is an Assistant Professor at the Vienna University of Technology (TU Wien) and a group leader at the the Complexity Science Hub Vienna (CSH). 
Her research mainly focuses on computational and network approaches to address societal challenges such as gender disparities in collaboration and citation networks~\cite{jadidi2018gender, kong2022influence}, visibility of minorities in social and technical systems~\cite{karimi2018homophily, espin2022inequality, oliveira2022group}, algorithmic biases~\cite{ntoutsi2020bias, espin2021explaining, ferrara2022link}, and sampling hard-to-reach groups~\cite{wagner2017sampling, espin2018towards}. 
Her research also touches upon the emergence of culture in Wikipedia~\cite{karimi2015mapping, samoilenko2016linguistic}, spreading of information and norms~\cite{kohne2020role}, and perception biases~\cite{lee2019homophily} by using mathematical models, digital traces and online experiments.

\section{Topic and relevance}

In this tutorial, we aim to cover two paradigms related to social network modeling and some applications.

\begin{enumerate}
    \item From social theories to models (120')
    \begin{enumerate}
        \item Social theories of edge formation
        \item Network properties and srtucture
        \item Network models
    \end{enumerate}
    
    \item From data to models (120')
    \begin{enumerate}
        \item Model fitting / inference
        \item Model selection
        \item Disentangling homophily and triadic closure
    \end{enumerate}
    
    \item Applications (70')
    \begin{enumerate}
        \item Biases in node sampling
        \item Inequalities in node ranking
    \end{enumerate}
    \item Challenges and open questions (20')
\end{enumerate}

\subsection{Social theories of edge formation}
Understanding how networks form is a key interest for ``The Web Conference" community.
For example, social scientists are frequently interested in studying relations between entities within social networks, e.g., how social friendship ties form between actors and explain them based on attributes such as  a person's gender, race, political affiliation or age in the network \cite{sampson1968novitiate}.
Similarly, the complex networks
community suggests a set of generative network models aiming at explaining the formation of edges focusing on the two core principles of \emph{popularity} and \emph{similarity} \cite{papadopoulos2012popularity}. 
Thus, a series of approaches to study edge formation have emerged including statistical tools \cite{krackhardt1988predicting,snijders1995use} and model-based approaches \cite{snijders2011statistical,papadopoulos2012popularity,karrer2011stochastic} specifically established in the physics and complex networks communities. 
Other disciplines such as the computer sciences, and political sciences use these tools to understand how co-authorship networks\cite{martin2013coauthorship} or online communities~\cite{adamic2005political} form or evolve.

In terms of similarity, many social networks demonstrate a property known as homophily, which is the tendency of individuals to associate with others who are similar to them, e.g., with respect to gender or ethnicity~\cite{mcpherson2001birds}.
Alternatively, individuals may also prefer to close triangles by connecting to people whom they already share a friend with~\cite{granovetter1973strength} which in turn can explain the emergence of communities~\cite{bianconi2014triadic}, high connectivity~\cite{newman2001clustering}, and induced homophily~\cite{asikainen2020cumulative}.
Furthermore, the class balance or distribution of individual attributes over the network is often uneven, with coexisting groups of different sizes, e.g., one ethnic group may dominate the other in size. 
Popularity, on the other hand, often refers to how well connected a node is in the network which in turn creates an advantage over poorly connected nodes. This is also known as the rich-get-richer or Matthew effect when new nodes attach preferentially to other nodes that are already well connected~\cite{barabasi1999emergence}. 
Many networks, including the World Wide Web, reflect such property by means of scale-free power-law degree distributions.

Here we will focus on the main mechanisms of edge formation namely homophily, triadic closure, node activity, and preferential attachment. Moreover, we will pay special attention to certain structural properties of networks such as class (im)balance, directed edges, and edge density.

\subsection{Network models}
In this section, we will review a set of well known network generator models. 
We will cover attributed graphs where nodes possess metadata information such as class membership, and edges are influenced by such information. The implementation of these models can be found in the \code{netin} python package.

\begin{enumerate}
    \item Attributed undirected graphs
    \begin{enumerate}
        \item Preferential attachment (PA)
        \item Preferential attachment with homophily (PAH)
        \item Preferential attachment with homophily and triadic closure (PATCH)
    \end{enumerate}
    \item Attributed directed graphs
    \begin{enumerate}
        \item Preferential attachment (DPA)
        \item Homophily (DH)
        \item Preferential attachment with homophily (DPAH)
    \end{enumerate}
\end{enumerate}

\subsection{Model selection and validation}
Identifying the model that best explains a given network remains an open challenge. 
First, we will show how to infer the hyper-parameters of each network model (e.g., homophily and triadic closure~\cite{peixoto2022disentangling}) given a real-world network. Then, we will learn how to use and interpret different approaches including AIC~\cite{williams2011probabilistic}, BIC~\cite{airoldi2008mixed}, MDL~\cite{peixoto2015model}, Bayes factors~\cite{hofman2008bayesian}, and likelihood ratios~\cite{yan2014model}, and highlight their strengths and limitations under specific tasks.

\subsection{Applications}
Here, we will demonstrate how to exploit network models to generate a wide range of synthetic networks to understand how certain algorithms are influenced by network structure and edge formation. The idea is to evaluate the outcomes of the following algorithms and see how they change while also changing the input network.

\subsubsection{Biases in node sampling.}
A range of network properties such as degree and betweenness centrality have been found to be sensitive to the choice of sampling methods~\cite{leskovec2006sampling,Lee2015,Wagner2017}.
These efforts have shown that network estimates become more inaccurate with lower sample coverage, but there is a wide variability of these effects across different measures, network structures and sampling errors.
In terms of benchmarking network sampling strategies, \cite{coscia2018benchmarking} shows that it is not enough to ask which method returns the most accurate sample (in terms of statistical properties); one also needs to consider API constraints and sampling budgets~\cite{espin2018towards, espin2021explaining}.

\subsubsection{Inequalities in node rankings.}
Previous studies have shown that homophily and group-size affect the visibility of minorities in centrality rankings~\cite{karimi2018homophily, fabbri2020effect, espin2022inequality}. 
In particular, such structural rankings may reduce, replicate and amplify the visibility of minorities in top ranks when majorities are homophilic, neutral and heterophilic, respectively. In other words, minorities are not always under-represented, they are just not well connected, and this can be shown by systematically varying the structure of synthetic networks~\cite{espin2022inequality}.
Here, we will also touch upon interventions on how to improve the visibility of minorities in degree rankings~\cite{neuhauser2022improving}.

\subsubsection{Biases in network inference.}
In recent years, there has been an increase of research focusing on mitigating bias \cite{raghavan2020mitigating,krasanakis2018adaptive} and guaranteeing individual and group fairness while preserving accuracy in classification algorithms \cite{dwork2018decoupled, binns2020apparent,kallus2019assessing,zafar2017fairnessbeyond}. 
While all this body of research focuses on fairness influenced by the attributes of the individuals, recent research proposes a new notion of fairness that is able to capture the relational structure of individuals \cite{farnadi2018fairness,zhang2020learning}.
An important aspect of \textit{explaining discrimination} \cite{mehrabi2019survey} via network structure is that we gain a better understanding of the direction of bias (i.e., why and when inference discriminates against certain groups of people)~\cite{espin2021explaining}.

\subsubsection{Inequalities in spreading dynamics}
Spreading processes may include simple and complex contagion mechanisms, different transmission rates within and across groups, and different seeding conditions. Here, we will study information access equality to demonstrate to what extent network structure influences a spreading process which in turn may affect the equality and efficiency of information access~\cite{wang2022information}.

\subsection{Challenges and open questions}
We will conclude by summarizing what we have learned, and by brainstorming future directions of what is still missing for producing realistic networks via synthetic data.

\section{Style, Duration, and Material}
This will be a 6-hour hybrid hands-on tutorial. We will provide ready to use jupyter notebooks with all necessary code, libraries, and settings. We will be using \code{python=3.9} and libraries such as:
\begin{enumerate}
    \item \code{networkx=2.8.8}
    \item \code{netin=1.0.7}
    \item \code{graph-tool=2.45}
    \item \code{matplotlib=3.6.0} 
    \item \code{numpy=1.23.4}
    \item \code{pandas=1.5.1}
    \item \code{jupyterlab=3.6}
\end{enumerate}

We will provide the slides of the tutorial beforehand, as well as code in the form of python scripts and notebooks. We will also use publicly available real-world networks~\cite{peixoto2020data}.
\textbf{All materials can be found here,}\footnote{\url{https://bit.ly/snma2023}} \textbf{and a video teaser of this tutorial here.}\footnote{\url{https://bit.ly/TutorialWWWTeaserENPK2023}}

\section{Previous editions}
This is the first time the organizers together have conceptualized and planned this tutorial. However, it will not be the first time they organize and teach network science to a broad audience.
\textit{Tiago Peixoto} has an extensive record in organizing workshops\footnote{\url{https://sinm.network/}}, and teaching at seminars and international schools on topics about data science, network science, and probabilistic and statistical methods for networks\footnote{\url{https://bayesforshs.sciencesconf.org/}}. 
\textit{Fariba Karimi} has given lectures and seminars on network science, theory, and dynamics to a broad audience including computer scientists and social scientists at the University of Koblenz-Landau and GESIS --- The Leibniz Institute for the Social Sciences.
\textit{Lisette Espin-Noboa} co-organized and co-lectured in 2020 a 4-day virtual hands-on seminar for social scientists on how to do network analysis in Python~\cite{espin2020seminar}. 
Additionally, Karimi and Espin-Noboa, co-organized a virtual satellite event at Networks 2021 where they invited a diverse group of researchers to talk about their research on network structure and social phenomena\footnote{\url{https://bit.ly/NetStructure}}.\section{Equipment}
We will require connection to the internet, a projector, and host permissions in Zoom for screen sharing, breakout rooms assignment, and remote access if necessary. Attendees may join the session online or in person using their own computers.\section{Organization details}
In case of unexpected events (e.g., restricted mobility, sickness, or bad internet connection) we will provide pre-recorded lectures of the entire tutorial. Moreover, all exercises will be given in advance as \code{python} scripts and Jupyter notebooks.

\bibliographystyle{ACM-Reference-Format}

\end{document}